\documentclass[showpacs,preprint,amsmath,superscriptaddress,prl]{revtex4}  
\usepackage[dvips]{graphicx}
\usepackage{epsfig}
\usepackage{color}
\usepackage{subfigure}
\usepackage{soul}
\usepackage{amssymb,amsmath,color,amsthm}

\begin{document} 
\title{
Finite-time scaling in local bifurcations
} 
\author{\'Alvaro Corral}
\affiliation{
Centre de Recerca Matem\`atica,
Edifici C, Campus Bellaterra,
E-08193 Barcelona, Spain
}
\affiliation{Departament de Matem\`atiques,
Facultat de Ci\`encies,
Universitat Aut\`onoma de Barcelona,
E-08193 Barcelona, Spain}
\affiliation{Barcelona Graduate School of Mathematics, Campus de Bellaterra, Edifici C, 08193 Bellaterra, Barcelona, Spain}
\affiliation{Complexity Science Hub Vienna,
Josefst\"adter Stra$\beta$e 39,
1080 Vienna,
Austria
}
\author{Josep Sardany\'es}
\affiliation{
Centre de Recerca Matem\`atica,
Edifici C, Campus Bellaterra,
E-08193 Barcelona, Spain
}
\affiliation{Barcelona Graduate School of Mathematics, Campus de Bellaterra, Edifici C, 08193 Bellaterra, Barcelona, Spain}
\author{Llu\'is Alsed\`a}
\affiliation{Barcelona Graduate School of Mathematics, Campus de Bellaterra, Edifici C, 08193 Bellaterra, Barcelona, Spain}
\affiliation{Departament de Matem\`atiques,
Facultat de Ci\`encies,
Universitat Aut\`onoma de Barcelona,
E-08193 Barcelona, Spain}
\begin{abstract} 
Finite-size scaling is a key tool in statistical physics, used to infer critical behavior in finite
systems. 
Here we use the analogous concept of finite-time scaling to describe the bifurcation
diagram at finite times in discrete dynamical systems. 
We analytically derive finite-time scaling
laws for two ubiquitous transitions given by the transcritical and the saddle-node bifurcation,
obtaining exact expressions for the critical exponents and scaling functions. One of the scaling
laws, corresponding to the distance of the dynamical variable to the attractor, turns out to be
universal. Our work establishes a new connection between thermodynamic phase transitions
and bifurcations in low-dimensional dynamical systems, 
and opens new avenues to identify the nature of dynamical shifts in systems for which only short time series are available.
\end{abstract} 

\date{\today}

\maketitle

\section*{Introduction}

Bifurcations separate qualitatively different 
dynamics in dynamical systems as one or more 
parameters are changed.
Bifurcations have been mathematically characterized in 
elastic-plastic materials \cite{Nielsen1993}, 
electronic circuits \cite{Kahan1999}, 
or in open quantum systems \cite{Ivanchenko2017}. 
Also, bifurcations have been theoretically described 
in population dynamics \cite{Rietkerk2004,%
Staver2011,Carpenter2011}, 
in socioecological systems \cite{May2008,Lade2013}, 
as well as in 
fixation of alleles in population genetics and computer virus propagation, to name a few examples 
\cite{Murray2002, Ott2002%
}. 
More important, bifurcations have been identified experimentally 
in physical \cite{Gil1991,Trickey1998,Das2007,Gomez2017},
chemical  \cite{Maselko1982,Strizhak1996}, 
and biological systems \cite{Dai2012,Gu2014}. 
The simplest cases of local bifurcations,
such as the transcritical and the saddle-node bifurcations,
only involve changes 
in the stability and existence of fixed points. 

Although, strictly speaking, 
attractors (such as 
stable fixed points) are only reached in the infinite-time limit,
some studies near local bifurcations
have focused on the dependence of the characteristic time needed to approach the  
attractor as a function of the 
distance of the bifurcation parameter to the bifurcation point. 
For example, for the transcritical bifurcation it is known that the transient time, 
$\tau$, diverges as a power law \cite{TEIXEIRA}, 
as $\tau \sim |\mu - \mu_c|^{-1}$, with
$\mu$ and $\mu_c$ being, respectively, the bifurcation parameter and the bifurcation point, while for the saddle-node bifurcation this time goes as 
$\tau \sim |\mu - \mu_c|^{-1/2}$ \cite{Strogatz_book} 
(see \cite{Trickey1998} for an experimental evidence of this power law in an electronic circuit). 

Thermodynamic phase transitions \cite{Stanley,Yeomans1992},
where an order parameter sudden changes its behavior as a response to
small changes in one or several control parameters,
can be considered 
as bifurcations. 
Three important peculiarities of thermodynamic phase transitions
within this picture
are that 
the order parameter has to be equal to zero in one of the phases or regimes,
that
the bifurcation does not arise (in principle) from a simple low-dimensional dynamical system
but from the cooperative effects of many-body interactions, 
and that at thermodynamic equilibrium there is no (macroscopic) dynamics at all.
Non-equilibrium phase transitions \cite{Marro_Dickman,Munoz_colloquium} 
are also bifurcations and share these characteristics, except the last one.
Particular interest has been paid to second-order phase transitions,
where the sudden change of the order parameter is nevertheless continuous
and associated to the existence of a critical point.

A key ingredient of second-order phase transitions is finite-size scaling \cite{Barber,Privman},
which describes how the sharpness of the transition emerges in the thermodynamic 
(infinite-system) limit.
For instance, if $m$ is magnetization (order parameter), 
$T$ temperature (control parameter), 
and $\ell$ system size, for zero applied field and close to the critical point 
the equation of state can be approximated as a finite-size scaling law,
\begin{equation}
m\simeq \frac 1 {\ell^{\beta/\nu}} g[\ell^{1/\nu} (T-T_c)],
\label{eq_state_scaling}
\end{equation}
with $T_c$ the critical temperature, 
$\beta$ and $\nu$ two critical exponents, and 
$g[y]$ a scaling function fulfilling 
$g[y] \propto (-y)^\beta$ for $y \rightarrow -\infty$
and
$g[y] \rightarrow 0$ for $y \rightarrow \infty$.

It has been recently shown that the Galton-Watson branching process
(a fundamental stochastic model for the growth and extinction of populations, 
nuclear reactions, and avalanche phenomena) can be understood as displaying
a second-order phase transition \cite{Corral_FontClos} with finite-size scaling  
\cite{GarciaMillan,Corral_garciamillan}.
In a similar spirit, in this article we show how bifurcations in one-dimensional 
discrete dynamical systems display ``finite-time scaling'', 
analogous to finite-size scaling with time playing the role of system size.
We analyze the transcritical and the saddle-node bifurcations for iterated maps and
find analytically well-defined scaling functions
that generalize the bifurcation diagrams for finite times.
The sharpness character
of each bifurcation is naturally recovered
in the infinite-time limit.
As a by-product, we derive 
the power-law divergence of the characteristic time $\tau$
when $\mu$ is kept constant, off of criticality \cite{TEIXEIRA,Strogatz_book}.



\section{Universal convergence to attractive fixed points}
Let us consider a one-dimensional 
discrete dynamical system, or
iterated map, 
$x_{n+1} =f(x_n),$
where $x$ is a real variable, 
$f(x)$ is a univariate function
(which will depend on some non-explicit parameters)
and $n$ being discrete time. 
Let us consider also that the map has an attractive (i.e., stable) fixed point at $x=q$,
for which $f(q)=q,$ and that $x_0$ belongs to the domain of attraction of the fixed point
(more conditions on $x_0$ later).
It is important to remember that attractiveness 
in discrete-time systems is characterized by $|f'(q)|<1$
(where the prime denotes derivative) \cite{Strogatz_book}.

We are interested in the behavior of $x_n=f^n (x_0)$ for large but finite $n$,
where $f^n (x_0)$ denotes the iterated application of the map $n$ times.
Naturally, for sufficient large $n$, 
$f^n (x_0)$ will be close to the attractive fixed point $q$
and we will be able to expand 
$f(f^n (x_0))$ around $q$, 
so,
\begin{eqnarray}
f^{n +1}(x_0) &=& f(f^n (x_0)) =q+M(f^n (x_0)-q) \nonumber \\ &+&C (f^n (x_0)-q)^2 
+\mathcal{O}(q-f^n (x_0))^3,
\label{taylor}
\end{eqnarray}
with $$M=f'(q) \mbox{ and } C=\frac{f''(q)} 2.$$
Rearranging and introducing the variable $c_{n+1}$, 
the inverse of the distance to the fixed point at iteration $n+1$,
we arrive to
$$
c_{n +1} = \frac 1 {q-f^{n +1}(x_0)}=\frac{c_n}M +\frac C{M^2}+
\mathcal{O}(q-f^n (x_0))
$$
(we may talk about a distance because, in practice, we calculate the difference
in such a way that it is always positive).
Iterating this transformation $\ell$ times we get
$$
c_{n +\ell} = \frac{c_n}{M^\ell} +\frac {C(1-M^\ell)}{M^{\ell+1}(1-M)},
$$
to the lowest order \cite{GarciaMillan}. 
Introducing the new variable $z=\ell(M-1)$, then,
for $\ell$ large one realizes that the second term in the sum
grows linearly with $\ell$ and overcomes the first one, and so, 
$
c_{n +\ell} \simeq {C \ell (e^z-1)} {e^{-z}}/z.
$
Next, considering $\ell$ much larger than $n$, 
so that $n+\ell \simeq \ell$, 
we get a scaling law for the dependence of the distance to the attractor
on $M$ and $\ell$,
\begin{equation}
q-f^\ell(x_0) = \frac 1 {c_\ell}\simeq
\frac 1 {C\ell} G(\ell(M-1)),
\label{scaling1}
\end{equation}
with scaling function
\begin{equation}
G(z) =\frac{z e ^z}{e^z-1}.
\label{scalingfunction}
\end{equation}
This result has also been obtained in Ref. \cite{GarciaMillan} 
for the Galton-Watson model, leading us to realize that this model is governed by a transcritical bifurcation (but restricted to $x_0 = 0$).


The scaling law (\ref{scaling1}) means that any attractor of a one-dimensional map
is approached in the same universal way,
as long as a Taylor expansion as the one in Eq. (\ref{taylor}) holds,
in particular if $f''(q) \ne 0$.
So we may talk about a ``universality class''.
The idea is that for different number of iterations $\ell$
one is able to find a value of $M$ (which depends on the parameters of $f(x)$)
for which $z=\ell(M-1)$ keeps constant and therefore the rescaled difference with respect the point $M=1$ is constant as well.
Note that, in order to have a finite $z$,  as $\ell$ is large, $M=f'(q)$ will be close to 1, 
so we will be close to a bifurcation point, corresponding to $M=1$
(where the attractive fixed point will lose its stability).
Due to this fact, in the scaling law we can replace $C$ by its value at the bifurcation point 
$C_*$, so, we write $C=C_*$ in Eq. (\ref{scaling1}).

In principle, the value of the initial value $x_0$ is not of fundamental importance, 
we could take for instance $x_1=f(x_0)$ as the initial condition instead, 
and we would get the same result just replacing $\ell$ by $\ell-1$.
For very large $\ell$ this difference plays no role ($\ell\simeq \ell-1$).
Therefore, as $\ell$ grows, 
the influence of the initial condition gets lost, 
as we can make $\ell$ as large as desired. 
But on the other hand, $x_0$ has to fulfill
$x_0<q$ if $C_*>0$ 
and $x_0 > q $ if $C_*<0$,
in the same way that all the iterations $x_n$
(i.e., all the iterations have to be on the same ``side'' of $q$).
The scaling law implies that plotting $[q-f^\ell(x_0)] C_* \ell$ versus 
$\ell (M-1)$ has to yield a data collapse of the curves corresponding to different 
values of $\ell$
onto the scaling function $G$.

\begin{figure*}
\center
\includegraphics[width=\textwidth]{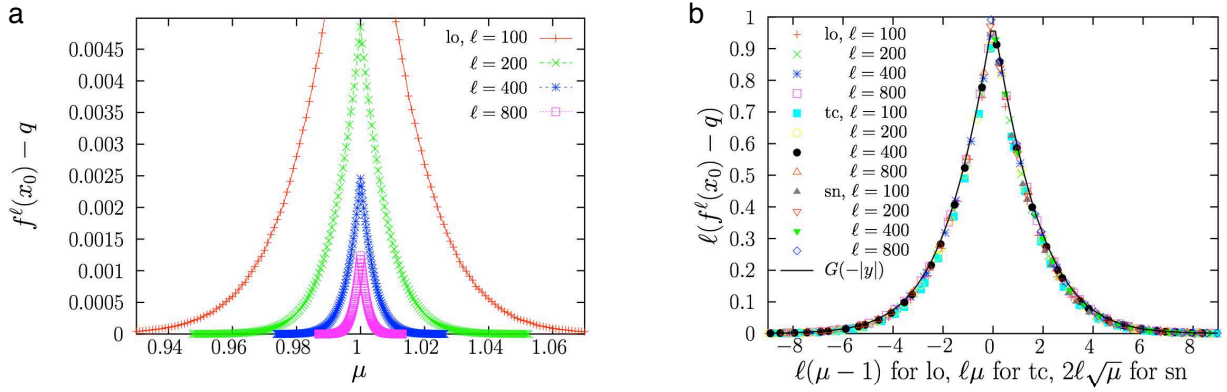}
\caption{
(a) Distance between the $\ell-$th iteration of the logistic map (lo) and
its attractor, as a function of the bifurcation parameter $\mu$,
for different values of $\ell$.
(b) The same data under rescaling (decreasing the density of points, for clarity sake),
together with data from the transcritical bifurcation in normal form (tc)
and the saddle-node bifurcation (sn).
The collapse of the curves into a single one validates the scaling law, 
Eq. (\ref{scaling1}), and its universal character.
The scaling function is in agreement with
$G(-|y|)$.
Note that the initial condition $x_0$ is taken uniformly randomly between 0.25 and 0.75, 
which is inside the range necessary for all the iterations to be above the fixed point.
This range is, below the bifurcation point, 
$ 0<x_0<1$ (lo),
$ 0<x_0<1+\mu$ (tc),
and, above,
$ 1-\mu^{-1}<x_0<\mu^{-1}$ (lo),
$\mu <x_0<1$ (tc),
$ \sqrt{\mu}<x_0<1-\sqrt{\mu}$ (sn).
}
\label{figone}
\end{figure*}

For example, 
for the logistic (lo) map \cite{Strogatz_book},
$f(x)=f_{lo}(x)=\mu x (1-x)$, 
a transcritical bifurcation takes place at $\mu=1$
and the attractor is at $q=0$ for $\mu \le 1$
and at $q=1-1/\mu$ for $\mu \ge 1$, 
which leads to $M_{lo}=f_{lo}'(q)=\mu$ for $\mu \le 1$
and $M_{lo}=2-\mu$ for $\mu \ge 1$, and also to $C_{lo *}=-1$. 
Therefore, $z=\ell(M-1)=-\ell|\mu-1|$ and
$
f_{lo}^\ell(x_0)-q 
\simeq  \ell^{-1} G(-\ell |\mu -1|),
$
for $x_0>q$.
Thus, in order to verify the collapse of the curves
onto the function $G$, 
one needs to represent $[f_{lo}^\ell(x_0)-q] \ell$ 
versus $-\ell |\mu -1|$,
or, if one wants to see separately the two regimes,
$ \mu \gtrless 1$, 
versus $y=\ell (\mu -1)$.
In the latter case the scaling function turns out to be $G(-|y|)$.
Figure \ref{figone}(b) shows 
precisely this;
the nearly perfect data collapse for large $\ell$ is 
the indication of the fulfillment of the finite-time scaling law.
For comparison, 
Fig. \ref{figone}(a) shows the same data with no rescaling 
(i.e., just 
the distance to the attractor as a function of the bifurcation parameter $\mu$).

{If one prefers the normal form of the transcritical (tc) bifurcation (in the discrete case), 
$f_{tc}(x)=(1+\mu)x-x^2$, 
then the bifurcation takes place at $\mu=0$
(with $q=0$ for $\mu \le 0$ and $q=\mu$ for $\mu\ge 0$). 
This leads to exactly the same behavior for $z=-\ell|\mu|$
(or for $y=\ell \mu$ in order to separate the two regimes, 
as shown overimposed in Fig. \ref{figone}(b), again with very good agreement).
}

For the saddle-node (sn) bifurcation
(also called fold or tangent bifurcation \cite{Kuznetsov_book}), 
in its normal form (discrete system), 
$f_{sn}(x)=\mu +x - x^2$, the attractor is at $q=\sqrt{\mu}$ 
(only for $\mu > 0$),
so the bifurcation is at $\mu=0$,
which leads to $M_{sn}=
1-2\sqrt{\mu}$ and $C_{sn *}=-1$.
The scaling law can be written as
\begin{equation}
f_{sn}^\ell(x_0) - \sqrt{\mu} \simeq \frac 1 \ell G(-2 \ell \sqrt{\mu}).
\label{scalingsn}
\end{equation}
To see the data collapse onto the function $G$
one must represent $[f_{sn}^\ell(x_0)-\sqrt{\mu}] \ell$ 
versus $z=-2 \ell \sqrt{\mu}$
(or versus $y=-z$ for clarity sake, as shown also in Fig. \ref{figone}(b)).
If one prefers a horizontal axis linear in $\mu$, 
one may define 
$z=-\sqrt{u}$, 
and then $f_{sn}^\ell(x_0)-\sqrt{\mu}\simeq F(4 \ell^2 \mu) / \ell$,
with a transformed scaling function
$
F(u)=G(-\sqrt{u})= {\sqrt{u}} / {(e^{\sqrt{u}}-1)},
$
and then use $u=-z^2=4 \ell^2 \mu$ for the horizontal axis
of the rescaled plot.

{Although the key idea of the finite-time scaling law, Eq. (\ref{scaling1}),
is to compare the solution of the system at ``corresponding'' values 
of $\ell$ and $\mu$ (such that $z$ is constant,
in a sort of law of corresponding states \cite{Stanley}),
the law can be used as well at fixed $\mu$.
%
At the bifurcation point ($\mu=\mu_c$, so $z=0$), 
we find that the distance to the attractor
decays hyperbolically, 
i.e., $|f^\ell(x_0)-q| =|C_* \ell|^{-1}$, 
as it is well known, see for instance Ref. \cite{TEIXEIRA}.
Out of the bifurcation point, for non-vanishing $\mu-\mu_c$ we have $z \rightarrow -\infty$
(as $\ell \rightarrow \infty$)
and then $G(z) \rightarrow e^{-z}$, 
which leads to $f^\ell(x_0)-q \simeq \ell^{-1} e^{-z} \simeq e^{-\ell /\tau}$,
where, from the expression for $z$,
we find that the characteristic time $\tau$ diverges as $\tau =1/|\mu-\mu_c|$
for the transcritical bifurcation (both in normal form and in the logistic form)
and as $\tau =1/(2\sqrt{\mu-\mu_c})$ for the saddle-node bifurcation 
(with $\mu_c=0$ in the normal form) \cite{Trickey1998}.
These laws, mentioned in the introduction, 
have been reported in the literature as scaling laws \cite{Strogatz_book}, 
but in order to avoid confusion we suggest to call them
power-law divergence laws.
Note that this sort of law arises
because $G(z)$ is asymptotically exponential;
in contrast, the equivalent of $G(z)$ in the equation of state of a magnetic system
in the thermodynamic limit is a power law, which leads to 
the Curie-Weiss law \cite{Christensen_Moloney}.

}

%

\section{Scaling law for 
the distance to the fixed point at bifurcation for the
iterated value $x_n$ in the  
transcritical bifurcation}
In some cases, 
the distance between $f^\ell(x_0)$ and some constant value of reference will
be of more interest than the distance to the attractive fixed point $q$,
as the value of $q$ may change with the bifurcation parameter.
For the transcritical bifurcation we have two fixed points, $q_0$ and $q_1$,
and they collide and interchange their character 
(attractive to repulsive, and vice versa) at the bifurcation point. 
Let us consider that $q_0$ is constant independently of the bifurcation parameter
(naturally, $q_1$ will not be constant), 
and that ``below'' the bifurcation point $q_0$ is attractive
and $q_1$ is repulsive, and vice versa ``above'' the bifurcation.
We will be interested in the distance between $q_0$ and $f^\ell(x_0)$, 
i.e., $q_0-f^\ell(x_0)$, 
which, below the bifurcation point 
corresponds to the quantity calculated previously in 
Eq. (\ref{scaling1}),
but not above. 
The reason is that, in there, $q$ was an attractor, 
but now $q_0$ can be attractive or repulsive.
Note that, without loss of generality, we can refer $q_0-f^\ell(x_0)$
as the distance of $f^\ell(x_0)$ to the ``origin''.

Following Ref. \cite{GarciaMillan}, we need a relation between both fixed points when we are close 
to the bifurcation point. 
As, in that case, $q_1 \simeq q_0$, we can expand
$f(q_1)$ around $q_0$, to get
$$
f(q_1) = q_1=q_0+M_0(q_1-q_0)+C_0(q_1-q_0)^2+\mathcal{O}(q_1-q_0)^3,
$$
which leads directly to 
\begin{equation}
M_0-1=C_0(q_0-q_1),
\label{laqueusamos}
\end{equation}
to the lowest order in $(q_1-q_0)$.
Naturally, $M_0=f'(q_0)$ and $C_0=f''(q_0)/2$.
We will also need a relation between $M_1=f'(q_1)$ and $M_0$.
Expanding $f'(q_1)$ around $q_0$,
$
f'(q_1)=M_1=M_0+2C_0(q_1-q_0)+\mathcal{O}(q_1-q_0)^2,
$
which, using Eq. (\ref{laqueusamos}), leads to
\begin{equation}
M_0-1=1-M_1,
\label{laotra}
\end{equation}
to the lowest order.

Now let us write $q_0-f^\ell(x_0)=q_0-q_1+q_1-f^\ell(x_0)$.
For $q_0-q_1$ we will apply Eq. (\ref{laqueusamos}), 
and for $q_1-f^\ell(x_0)$ we can apply 
Eq. (\ref{scaling1}), as $q_1$ is of attractive nature ``above'' the bifurcation point;
then
$$
q_0-f^\ell(x_0)\simeq
\frac{M_0-1}{C_0}+\frac 1 {C_1\ell} G(\ell (M_1-1))
$$
(with $C_1=f''(q_1)/2$),
and defining $y=\ell(M_0-1)$ we get (with the form of the scaling function, Eq. (\ref{scalingfunction})),
$$
q_0-f^\ell(x_0)\simeq
\frac y {C_0\ell} +\frac 1 {C_1\ell} \left(\frac{ze^z}{e^z-1}\right).
$$
Using Eq. (\ref{laotra}) one realizes that
$z=\ell(M_1-1)=-\ell(M_0-1)=-y$
(so, the $y$ introduced here is the same $y$ introduced above), 
and therefore,
$$
q_0-f^\ell(x_0)\simeq
\frac 1 {C_* \ell}\left( y +\frac{-ye^{-y}}{e^{-y}-1}\right)=
\frac 1 {C_* \ell} \frac{ye^y}{e^y-1},
$$
where we have used also that $C_1=C_0=C_*$, to the lowest order,
with $C_*$ the value at the bifurcation point.
Therefore, 
we obtain the same scaling law as in the previous section:
\begin{equation}
q_0-f^\ell(x_0)\simeq \frac 1 {C_*\ell} G(y), 
\label{scalinglawguena}
\end{equation}
with the same scaling function $G(y)$ as in Eq. (\ref{scalingfunction}),
although the rescaled variable $y$ is different here ($y\ne z$, in general).
This is possible thanks to the property
$y+G(-y)=G(y)$
that the scaling function verifies.
Note that the scaling law (\ref{eq_state_scaling})
has the same form as the finite-time scaling (\ref{scalinglawguena})
and we can identify $\beta=\nu=1$.

Note also that we can identify $M_0=f'(q_0)$
with a bifurcation parameter, 
as it is $M_0<1$ ``below'' the bifurcation point ($M_0=1$)
and $M_0 > 1$ ``above''
($M$ defined in the previous section cannot be a bifurcation parameter
as it is never above 1, due to the fact that it is defined with respect the attractive fixed point). 

\begin{figure*}
\center
\includegraphics[width=\textwidth]{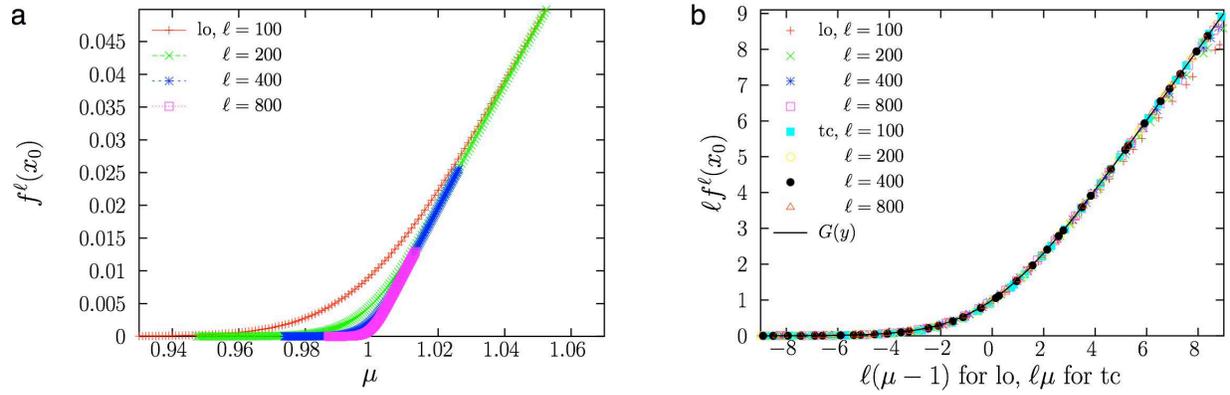}
\caption{
(a) $\ell-$th iteration of the logistic map as a function of the bifurcation parameter $\mu$, for different values of $\ell$.
Same initial conditions as in previous figure.
(b) Same data under rescaling (decreasing density of points), plus analogous data coming
from the transcritical bifurcation in normal form.
The data collapse shows the validity of the scaling law, Eq. (\ref{scalinglawguena}),
with scaling function $G(y)$ from Eq. (\ref{scalingfunction}).
}
\label{figtwo}
\end{figure*}

For the transcritical bifurcation of the logistic map we identify
$q_0=0$ and $M_0=\mu$, so $y=\ell(\mu-1)$.
For the normal form of the transcritical bifurcation,
$q_0=0$ but $M_0=\mu+1$, so $y=\ell \mu$.
Consequently, 
Fig. \ref{figtwo}(a) shows $f^\ell(x_0)$ (the distance to $q_0=0$)
as a function of $\mu$, for the logistic map and different $\ell$,
whereas Fig. \ref{figtwo}(b) shows the same results under the 
corresponding rescaling, together with analogous results 
for the normal form of the transcritical bifurcation.
The data collapse supports the validity of the scaling law (\ref{scalinglawguena})
with scaling function given by Eq. (\ref{scalingfunction}).


\section{Scaling law for the iterated value $x_n$ in the saddle-node bifurcation}
\begin{figure*}
\center
\includegraphics[width=\textwidth]{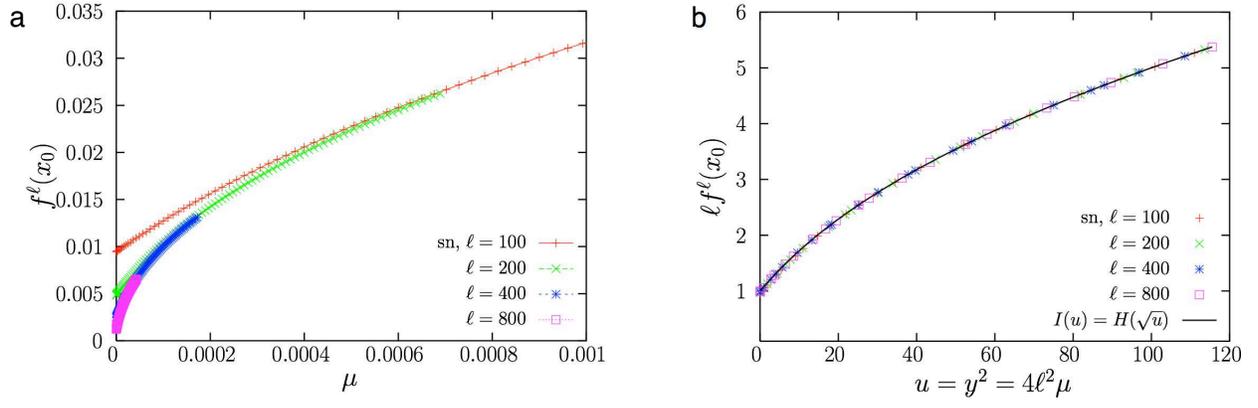}
\caption{
(a) Same as Fig. \ref{figtwo}(a) but for the saddle-node bifurcation in normal form.
(b) Rescaling of the same data (with decreased density of points).
The data collapse supports the scaling law and the scaling function $I(u)$ given by 
Eq. (\ref{ultimoscaling}).
}
\label{figthree}
\end{figure*}

Coming back to the saddle-node bifurcation, 
from Eq. (\ref{scalingsn}) we can isolate the $\ell-$th iterate to get, 
$$
f^\ell(x_0) \simeq \frac 1 \ell \left[
\frac{2\ell \sqrt{\mu}}2 +  G(-2\ell \sqrt{\mu}) \right] = \frac 1 \ell H(y)
$$
with $y=-z=2\ell \sqrt{\mu}$ and
$
H(y)={y(e^y+1)}{(e^y-1)^{-1}/2}.
$
Therefore, the representation of $\ell f^\ell(x_0) $ versus $2\ell \sqrt{\mu}$
unveils the shape of the scaling function $H$.
In terms of $u=y^2=4 \ell^2 \mu$,
\begin{equation}
f^\ell(x_0) \simeq
\frac 1 \ell I({u}),
\phantom{x}
{\rm{with}} \phantom{x} 
I(u)=
H(\sqrt{u})= \frac{\sqrt{u}} 2\frac{(e^{\sqrt{u}}+1)}{(e^{\sqrt{u}}-1)},
\label{ultimoscaling}
\end{equation}
and so, $\ell f^\ell(x_0) $ against $4 \ell^2 \mu$ leads to the collapse of the data
onto the scaling function $I(u)$, 
as shown in Fig. \ref{figthree}.
Comparison with the finite-size scaling law (\ref{eq_state_scaling})
allows one to establish $\beta=\nu=1/2$ for this bifurcation
(and bifurcation parameter $\mu$, not $\sqrt{\mu}$).

\section{Conclusions}
By means of scaling laws,
we have made clear an analogy between bifurcations
and phase transitions, with
a direct correspondence between, 
on the one hand, the bifurcation parameter, 
the bifurcation point, and the finite-time solution $f^\ell(x_0)$, 
and, on the other hand, 
the control parameter, the critical point, 
and the finite-size order parameter.  
However, 
in phase transitions, the sharp change of the order parameter
at the critical point arises in the limit of infinite system size;
in contrast, in bifurcations, the sharpness at the bifurcation point 
shows up in the infinite-time limit, $\ell \rightarrow \infty$.
So, finite-size scaling in one case corresponds to 
finite-time scaling in the other.

In addition,
we have also been able to derive the power-law divergence of the transient time
to reach the attractor off of criticality 
\cite{Trickey1998,Strogatz_book,TEIXEIRA}, 
and
also conclude that 
the results of Ref. \cite{GarciaMillan} can be directly understood from the
transcritical bifurcation underlying the Galton-Watson branching process.
Moreover, by using numerical simulations we have tested that 
the finite-time scaling laws also hold for 
dynamical systems continuous in time, 
as well as for the pitchfork bifurcation in discrete time (although with different exponents and scaling function in this case).
Let us mention that the use of the finite-time scaling concept by other authors
does not correspond with ours. For instance, although Ref. \cite{Agoritsas}
presents a scaling law for finite times, the corresponding exponent $\nu$ there 
turns to be negative, which is not in agreement with the genuine finite-size scaling
around a critical point.

Our results may also allow to identify the nature of bifurcations in systems for which information is limited to short transients, such as in ecological systems. 
In this way,
the scaling relations established in this article could be used as warning signals \cite{Scheffer2009} to anticipate the nature
of collapses or changes in ecosystems \cite{Staver2011,Carpenter2011,Scheffer2001,Scheffer2003,Scheffer2009} (due to, e.g., transcritical or saddle-node bifurcations) and in other dynamical suffering dynamical shifts.

\bibliographystyle{unsrt}

\section*{Acknowledgements}
We have received funding from ``La Caixa" Foundation and
through the ``Mar\'{\i}a de Maeztu'' Programme for Units of Excellence in R\&D (MDM-2014-0445), as well as from projects
FIS2015-71851-P and MTM2014-52209-C2-1-P from the Spanish 
MINECO, from 2014SGR-1307 (AGAUR),
and from the CERCA Programme of the Generalitat de Catalunya.

\section*{Author contributions statement}
All authors analysed and discussed the results. All authors reviewed the manuscript.

\section*{Additional information}

\textbf{Competing financial interests}. The authors declare no competing interests.

\newpage
\newpage
\clearpage

\newpage
\clearpage

\end{document}